\documentclass[12pt,english,preprint]{emulateapj}
\usepackage[latin9]{inputenc}                        
\makeatletter

\usepackage{epsfig}\usepackage{color}\usepackage{txfonts}\bibpunct{(}{)}{;}{a}{}{,} 

\newcommand{\omc}{$\omega$ Cen }
\newcommand{\omcv}{$\omega$ Cen}

\shorttitle{Evolution of stellar disks in GCs}\shortauthors{Mastrobuono-Battisti \& Perets}

\makeatother

\begin{document}

\title{Evolution of second generation stars in stellar disks of globular and nuclear clusters:\\
 $\omega$ Centauri as a test case}

\author{Alessandra Mastrobuono-Battisti\altaffilmark{1,2} \& Hagai B. Perets\altaffilmark{1}}

\altaffiltext{1}{Physics Department, Technion - Israel Institute of Technology, Haifa,
Israel 32000 }
\altaffiltext{2}{INAF - Osservatorio Astrofisico di Arcetri, Largo E. Fermi, 5, 50125 Firenze, Italy}
\begin{abstract}
Globular clusters (GCs) and many nuclear clusters (NCs) show evidence for hosting multiple
generations of stellar populations. Younger stellar populations in NCs appear to reside 
in disk like structures, including
the nuclear cluster in our own Galactic center as well as in M31.
Kinematic studies of the anomalous globular cluster $\omega$ Centauri,
thought to possibly be a former dwarf galaxy (or a galactic nucleus),
show evidence for its hosting of a central, kinematically cold disk
component. These observations suggest that formation of second (or
multiple) generation stars may occur in flattened disk like structures.
Here we use detailed N-body simulations to explore the possible evolution
of such stellar disks, embedded in globular clusters. We follow the long term evolution 
of a disk like structure similar to that observed in $\omega$ Centauri and study its properties. 
We find that a stellar-disk like origin for second generation stellar populations
can leave behind significant kinematic signatures in properties of the clusters, including an anisotropic 
distribution, and lower velocity dispersions, which can be used to constrain the 
origin of second generations stars and their dynamical evolution. 
\end{abstract}


\section{Introduction}\label{Intro}
Nuclear star clusters (NCs), 
stellar systems similar to globular clusters (GCs), have been found in $\sim 75$\% of the late type spiral galaxies and dwarf
ellipticals \citep{Bo04,Wa05,Co06,Bok10}.
They contain several distinct stellar populations, characterized by different ages and chemical properties. Among these groups of stars, 
the youngest ones often form centrally concentrated stellar disks \citep{VDB94, Seth06, Se08}. 
Such stellar structures are also found in the central pc of the Galaxy and in M31. Their populations of massive young ($6\pm 2$~Myr and $\sim200$~Myr, respectively)  stars are arranged in one (or more) distinct stellar disks \citep{T95,Ge00,Le03,Be05,Lu09,Ba10,Ge10}. 

Similarly, and in contrast with previous paradigm, GCs are not composed of a simple stellar population born at the same time from the same gas. Rather, stars in GCs show significant inhomogeneities in the abundance of light elements and their color magnitude diagrams show several isochrones; thus they consist of multiple stellar generations (MSGs) whose age spread is estimated
to extend up to a few hundred Myrs \citep[see e.g.][]{GSC04,GCB12}. 

The origin of these MSGs is still debated; in one of the models the low velocity gas ejected by  primordial asymptotic giant branch (AGB) stars is retained inside the cluster, concentrates at its center and there fragments forming  new stars \citep[see][and reference therein]{GCB12}. 
In this scenario, the initial distribution of second generation stars strongly depends on the 
initial profile of the collected gas (a somewhat similar model was suggested for the origin of M31 young stellar disk by \citealt{Ch07}).
For instance, if  the gaseous material has some angular momentum, it could produce a flattened structure instead of the commonly assumed 
spherical configuration \citep{Be11,Be10, MBP13}, in analogy with what is observed in NCs. 
This would influence significantly the later dynamical evolution and mixing of older and younger stars, as we show in the following.

One of first clusters where MSGs have been found is  $\omega$ Centauri (NGC 5139, \omcv, \citealt{JP10,Ma11}).
With mass estimates between $2\times10^{6}$ and $5\times10^{6}$~M$_{\odot}$
\citep[see][]{Mey95,VdV06}, \omc 
is the most luminous and massive Galactic GC.
It is characterized
by a flattened structure
\citep{Ge83,VdV06} and by a wide range of metallicities
\citep[e.g.][]{Bu78,NDC95}. Its MSGs are characterized by different spatial distributions \citep[e.g.][]{Fr75,Lee99,HW00,HR00,Pa00,Fe02,Be04,St04,So05, Be09}.
Dynamically, \omcv's orbit is peculiar, being tightly bound to the Galactic potential, retrograde and highly eccentric \citep{Di99}. 
Moreover, it is one of the favorite candidates for hosting an intermediate-mass black hole,
whose mass is estimated to be in the range $13,000\leq M_{\bullet}\leq50,000$~M$_{\odot}$
\citep{No08,vdM10,Mi10,Ja12}.
Given these features \omc is thought to be the remnant nucleus of a satellite dwarf galaxy that merged with the Milky Way
\citep{Fr93,Di99,HW00,BF03, Bo08}. \\

Recently \citet{VdV06} used an extension of the \citet{Sc79} method
to build axisymmetric dynamical models for \omcv, using anisotropic
velocity distributions. 
Their best fitting model shows a highly
isotropic inner region, which becomes increasingly anisotropic
outside of $10$~arcmin.\\
The nucleus of \omc contains a kinematically cold disk
component, with a radius between $1$ and $3$~arcmin, an average flattening
of $\sim0.60$,  and
contributing about $\sim4\%$ to the total mass of the cluster.

The presence of this disk could be closely related to the existence of
MSGs in \omc and inside  other GCs \citep{Be11, Be10}. 
In particular, second generation stars may form in a disk, and leave behind 
kinematic signatures of their former configuration potentially observable even after long term dynamical evolution. These signatures may allow us to infer information about their initial
distribution, and provide us with constraints on their origin.\\

In this paper we study the long term evolution of \omcv-like central disk by means of N-body simulations and explore its structure and potentially observable signatures.
This test case is used as a pilot to reveal the possible kinematic signatures of second-generation stellar disks and point to possible clues for the initial presence of a second generation disk in GCs and NCs.

The paper is organized as follows. In Section \ref{sec:NBS} we describe our choice of the initial conditions, the simulations we run and the code that we use. Then, in Section
\ref{sec:res}, we show the results obtained. These results and their implications
are discussed and summarized in Section \ref{sec:disc}.
\begin{deluxetable}{ccccccc}
\tablecolumns{7}
\tabletypesize{\small}
\tablewidth{0pc} 
\tablecaption{ \label{tabICs} Initial conditions for the simulation runs. }
\tablehead{ 
\colhead{Label}  & \colhead{$N$} & \colhead{$m_{*,1}$} & \colhead{$m_{*,2}$} & \colhead{$M_{disk}$ } & \colhead{$r_d$ } & \colhead{potential} \\
 & & \footnotesize{M$_\odot$} & \footnotesize{M$_\odot$} & \footnotesize{M$_\odot$} &\footnotesize{pc}  & }
\startdata
S1,2,3 & 4000 & 25 & 0 &$10^5$ & 4.5& isolated \\ 
S4 & 4000 & 25 & 0 & $10^5$ & 4.5 & tides\\
S5 & 10000 & 25 & 12.5 & $1.5\times 10^5$ & 4.5 & isolated \\
S6 & 2500 &  100 & 10 &  $10^5$ & 4.5 & isolated \\
S7 & 8000 & 12.5 & 0 & $10^5$ & 4.5 & isolated 
\enddata
\tablecomments{S1, S2 and S3 are three different realizations of the same initial conditions; in S1 we used a softening length one order of magnitude smaller that in S2 and S3.
S4 is similar to S1, but includes the effects of the Galactic tide, realized as an external point mass potential.
S5 and S6 realize disks with two different mass classes, each with a different mass ratio, while conserving the cluster mass. S7 is similar to S1 but models a larger number of lower mass stars, while conserving the cluster mass.}
\end{deluxetable}
%

\section{The N-body simulation} \label{sec:NBS}

\subsection{The simulation}

We studied the dynamical evolution of an \omcv-like central disk structure by means 
of N-body simulations. These simulations have been done using
$\phi$Grape \citep{Harfst}, a high precision, collisional, direct
summation code originally configured to run on computer clusters accelerated
by GRAPE boards \citep{Ma98}. This code has been recently adapted
to run on GPU equipped machines by means of Sapporo, a CUDA library
that calculates force on the GPUs mimicking the behavior of GPAPEs
\citep{Ga09}. \\
The code implements a fourth-order Hermite integrator with a predictor-corrector
scheme and hierarchical time stepping. The time-step parameter $\eta$
and the smoothing length $\epsilon$ are the parameters that controls
the accuracy of the code. 
In direct summation N-body simulation it is often necessary to use a softening parameter, $\epsilon$, in order to accelerate the calculations, and circumvent the short integration time steps required during close 
encounters between particles \citep{BT08}. 
Our use of a softening radius is due to computational expense constraints and is not generally desirable in simulations of a collisional system, and an appropriate softening length has to be chosen as not to detracts from the simulation correspondence to realistic systems. To verify that the use of a specific softening length does not significantly affect our results we compared both higher and lower resolution simulations ($\epsilon=5\times10^{-3}$~pc and $\epsilon=5\times10^{-2}$~pc). We find that the different simulations are compatible, and do not show any significant difference once the relaxation times are correctly rescaled.

Indeed, the effect of $\epsilon$ on the two-body relaxation process \citep[see e.g.][]{Th98,A01}
has to be taken into account when evaluating the 
relaxation time of the system, since its presence essentially increases the
minimum impact parameter in the Coulomb logarithm expression (see Section \ref{ICs}).
In our main simulation, we set $\eta=0.02$
and $\epsilon=5\times10^{-3}$~pc. As mentioned above, we also run several, faster, simulations with 
$\epsilon=5\times10^{-2}$~pc to test our results, and found that increasing the value of the softening length
does not affect the evolution of the system significantly.
Both the adopted values for the softening length are smaller than the core radius of the 
cluster and the mutual average distance between particles.
The choice made for the parameters $\eta$ and $\epsilon$ is then justified by the 
balance between the calculation time and the energy conservation.
The energy fractional variation at the end of each simulation, is indeed less than 
$10^{-4}$. 
The code has been run on the cluster ``Tamnun'' at the Technion.

\begin{figure}[!t]
\begin{center}
\includegraphics[trim=1.18cm 1.65cm 1.1cm 1.cm, clip=true,width=0.466\textwidth, angle=0]{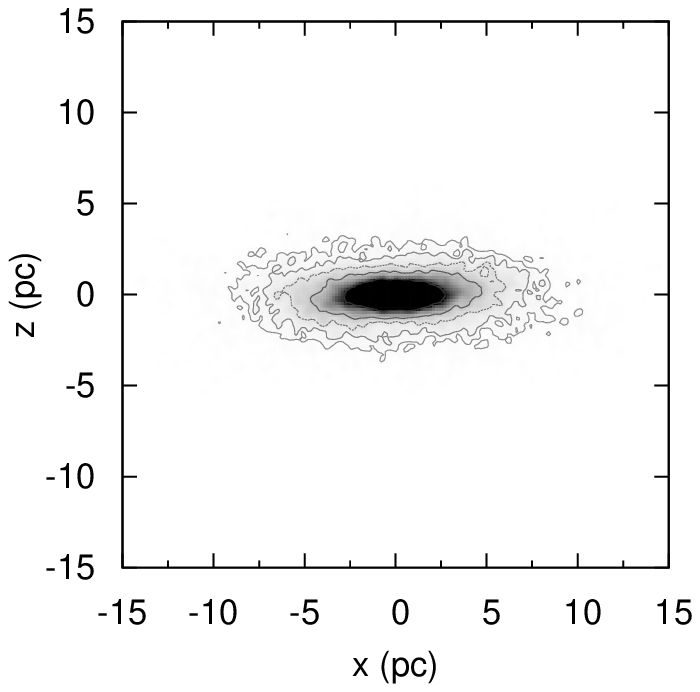}\\
\includegraphics[trim=1.18cm 1.7cm 1.1cm 1.cm, clip=true,width=0.466\textwidth, angle=0]{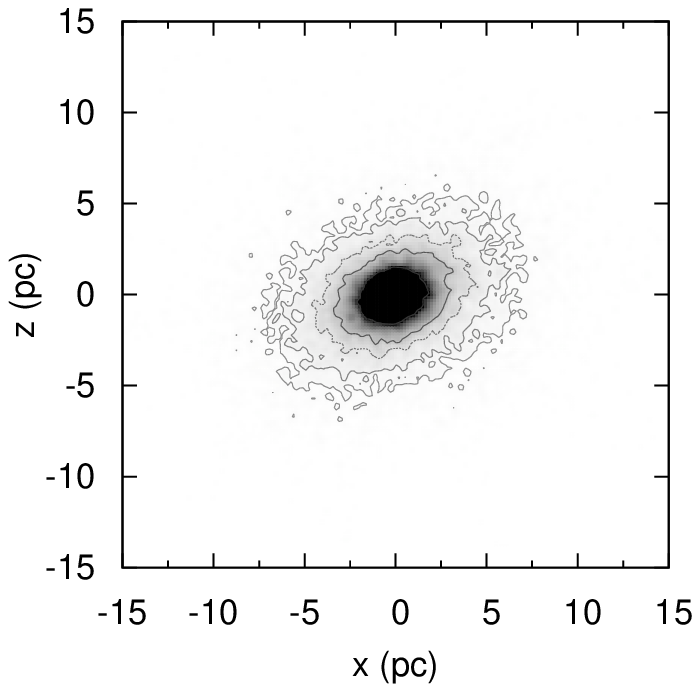}\\
\includegraphics[trim=1.18cm 1.7cm 1.1cm 1.cm, clip=true,width=0.466\textwidth, angle=0]{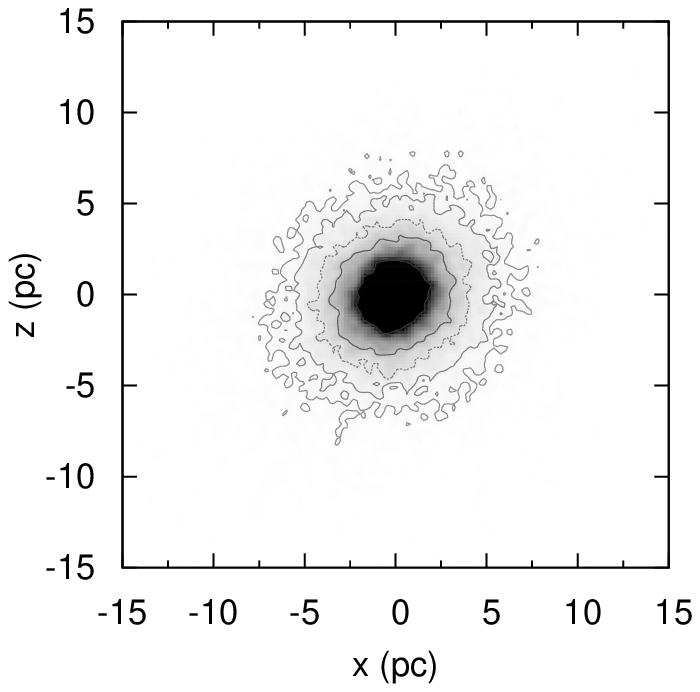}\\
\includegraphics[trim=1.14cm 0.4cm 1.1cm 1.cm, clip=true,width=0.469\textwidth, angle=0]{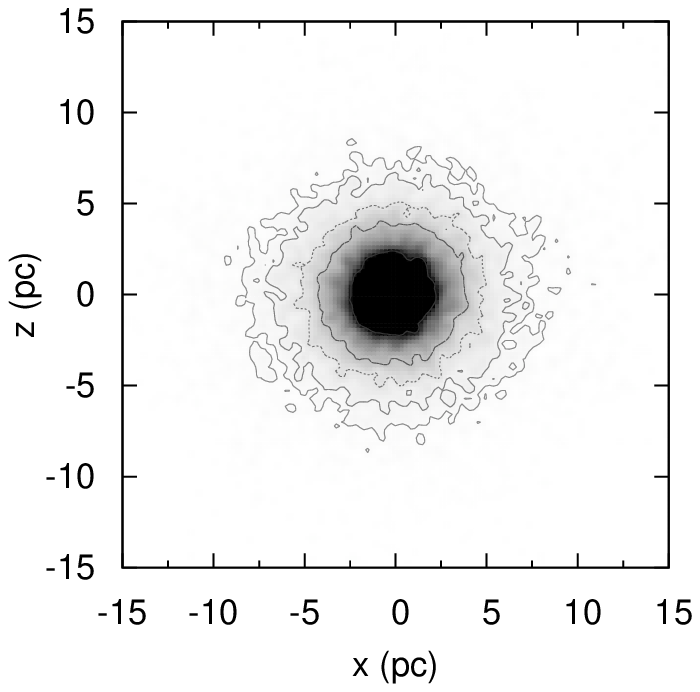}
\caption{Projected isodensity contours plot of the disk on the X-Z plane, at the beginning of the simulation,
and going downward, after 2, 5 and 12~Gyr. }
\label{fig:isoc}
\end{center}
\end{figure}

\subsection{Initial conditions}\label{ICs}

The initial conditions for our simulations were set based on the available observational
data. Since the relaxation time of \omc is longer than a Hubble
time \citep{Ha10}, mass segregation is not expected to have had
significant effects on the system. Its structure is therefore modeled as
a single-mass, spherical and isotropic \citet{K66} cluster. To build
our model we used the parameters that best fit the cluster observe
brightness profile: we set
$r_c=4.6$~pc as the core radius,
$W_{0}=5.5$ \citep{Me87} and
$2.5\times10^{6}$~M$_{\odot}$ as the total mass of the GC \citep[see][]{VdV06}. The N-body representation of the spherical component
consists of $N=100,000$ particles. 

In the main simulation we ran (S1, see Table \ref{tabICs}) the disk properties are similar to those observed in \omcv. The disk mass is $10^{5}$~M$_{\odot}$; it has a radial 
extension,  of about $4.5$~pc and the velocity dispersion of the disk stars is similar to that observed. 
 More specifically, these initial conditions have been produced by evolving a flat uniform disk, with $N_{d}=4000$ single mass particles 
orbiting counter clockwise in the X-Y plane. The initial velocity of the particles was taken to be 
equal to the circular velocity at the radius of the given particle in a Plummer sphere with parameters similar to those of the King model
adopted for the cluster. The initial radius of the flat disk is $8$~pc.
This configuration is unstable and it rapidly (only 1\% of the lifetime of the cluster) evolves to a quasi-stable state, where it becomes more centrally concentrated,
with more than the 80\% of the mass contained inside $\sim 4.5$~pc, and the rest extending to larger distances. The quasi-stable disk configuration obtained though this method is in good agreement with observations (compare with Figure 20 in \citealt{VdV06}); we therefore use this configuration as the initial conditions for our long term simulation of the disk evolution, i.e. at time $t=0$.
Note that the density profile of the disk after this initial setting (both on the X-Y and X-Z planes) is well 
fitted by two exponentially cut broken power laws, with a flat core of radius $\sim0.2$~pc.\\

Given the large computation time required for such large N-body simulations, our modeled system make use of less numerous but more massive particles to model the cluster, as to provide a realistic cluster mass.
Naturally, the evolutionary time scales are affected by this choice.
However, the N-body simulation results can be 
approximately scaled with $N$ using 
the method described by \citet{AH98}, as to produce realistic results. 
Using appropriate scaling provides the correct account of the relaxation times, as follows
\begin{equation}
t=t^*\frac{t_{rx}}{t_{rx}^*},
\end{equation}
where $t$ and $t^*$ are respectively the time in the real and in the simulated
systems, $t_{rx}$ is the relaxation time in the real cluster, and $t_{rx}^*$
is the corresponding quantity for the N-body model.
Following \citet{Spi87} the local relaxation time of a stellar system is given by
\begin{equation}
t_{rx} = \frac{0.065\sigma^3}{\rho m G^2\ln\Lambda}, \label{eq:relt}
\end{equation}
where $m$ is the stellar mass, $\rho$ is the mass density, and $\sigma$ is the
local velocity dispersion of the system. Then, we have
\begin{equation}
t=t^*\frac{m^*\ln\Lambda^*}{m\ln\Lambda},
\end{equation}
where the apex $^*$ refers to the model.
Using the mass function of \omc given by \cite{So07}
we obtained $m=0.48$~M$_\odot$ as the average mass of the stars in the cluster,
while for the mock system we have $m^*=25$~M$_\odot$.
Beside the rescaling with the particle number, we have to take into
account the introduction of a softened potential.
The value of the Coulomb logarithm is indeed affected by the introduction of
the softening length \citep[e.g.][]{A01}, since the value of the minimum 
softening length is artificially increased and so the time scales approximately as
\begin{equation}\label{eq:resc}
t=t^*\frac{m^*\ln(r_t/\epsilon)}{m\ln(N)}, 
\end{equation}
where we assumed the size of the cluster, namely its tidal radius $r_t$, as the maximum impact parameter and the deflection distance at $90^\circ$, i.e. $b_{90}\simeq Gm/v^2 \simeq r_t/N$ as the minimum impact parameter \citep[e.g.][]{A01, BT08}.

We assume the cluster has an initially isotropic distribution. Note that the currently observed cluster shows some anisotropies in its outer regions. However, as we briefly discuss later on, these might arise from the coupled evolution of the disk+cluster system, and might not be an inherent primordial property of the cluster.
\begin{figure}[!t]
\begin{center}
\includegraphics[width=0.4\textwidth]{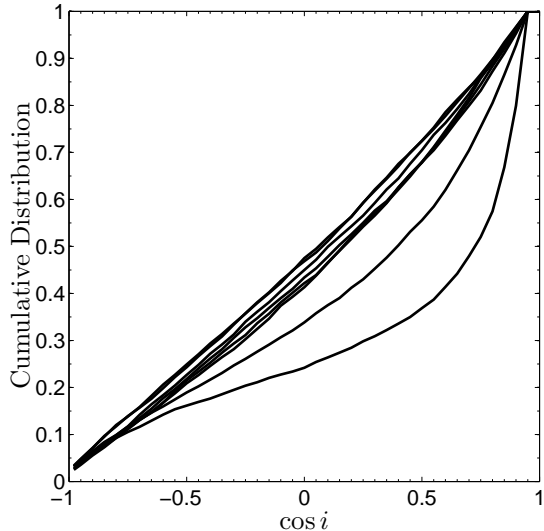}

\caption{
Cumulative distribution of inclination of stars in the disk at different, equally spaced, times.
Going upwards the curves correspond to 0, 1.5, 3, 4.5, 6, 7.5, 9, 10.5 and 12~Gyr. 
}
\label{fig:cosi}
\end{center}
\end{figure}
\begin{figure}[!t]
\begin{center}
\includegraphics[width=0.4\textwidth]{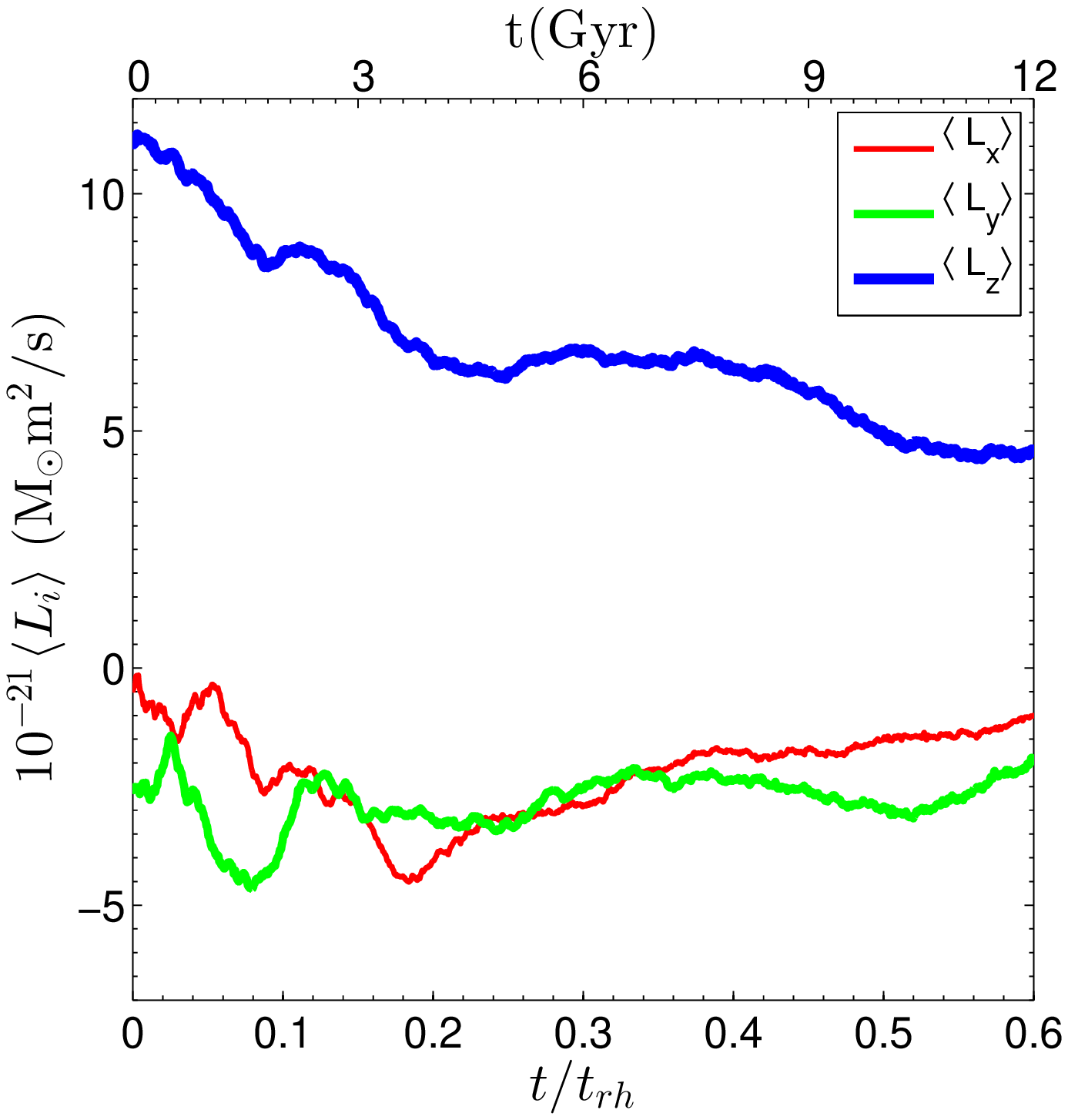}\\ 
\vspace{0.3cm}
\vspace{-0.2cm}\includegraphics[width=0.4\textwidth]{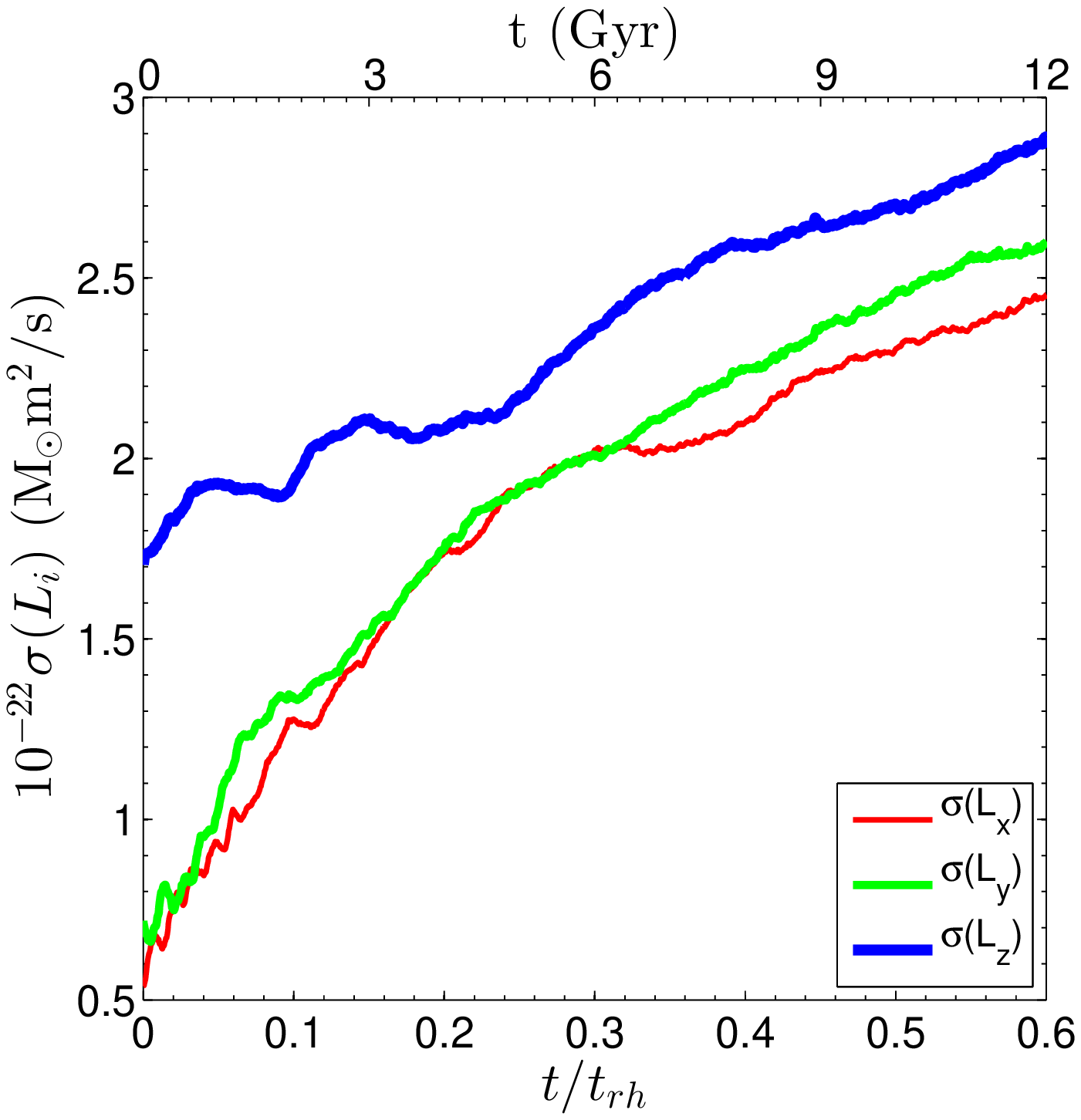}
\caption{The average angular momentum per particle in the disk. Upper panel: The evolution of the three components of the  average angular momentum of the disk particles. The $L_x$ and $L_y$ components lay on the initial plan of the disk, while $L_z$ is initially perpendicular to the disk. The disk loses angular momentum that is redistributed to the stars in the cluster. Bottom panel: The dispersion of the three components of the angular momentum.  The time is expressed both in Gyr and in terms of the half mass relaxation time of \omcv, evaluated using Equation (2.63) of \citet{Spi87}.}
\label{fig:Lev}
\end{center}
\end{figure}
In order to exclude spurious effects that might arise from specific assumptions made in our modeling, 
we ran another set of test simulations with different initial 
parameters for the disk (see Table \ref{tabICs}). 
In particular, we ran two additional realizations of the same disk+cluster system in order to verify that the results do not depend on the particular realization of the initial conditions (S2 and S3, see Table \ref{tabICs}). 
Due to computational limitations, these simulations were run using a softening length of $5\times10^{-2}$~pc, which is larger than the one 
used in S1. Nevertheless, after the rescaling given by Equation (\ref{eq:resc}), S1 and S2,3 provide compatible results.
We also tested the effects of the Galactic tide, by putting the cluster on a circular orbit around a point mass potential as massive as the Galaxy within \omcv's current position (S4). In every cases we find that the disk evolves 
in the same way as in the case of the first model (S1), and we therefore conclude that the presence of the Galactic tide and, the specific realization of the initial conditions and the chosen softening length do not introduce any significant effects. Later on we briefly discuss the slight effects due to the Galactic tide on the anisotropy of cluster external regions. 
\begin{figure}[!t]
\begin{center}
\includegraphics[width=0.41\textwidth]{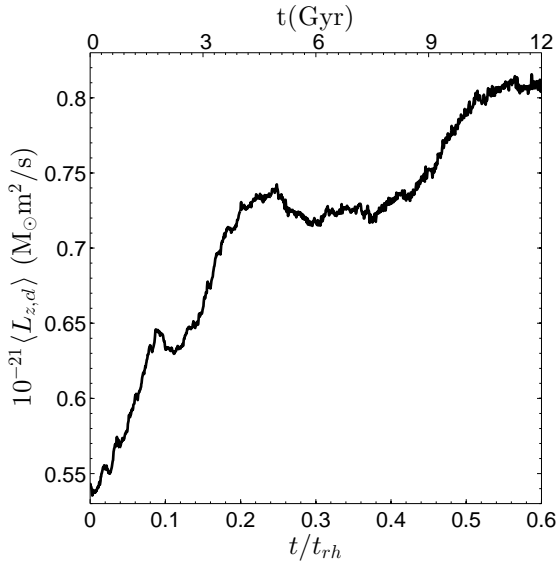}
\caption{The time evolution of the z-component of the  average angular momentum per 
particle in the initially spherical component
of our system.}
\label{fig:amz}
\end{center}
\end{figure}
In order to exclude any major effects in the disk evolution that might be introduced by mass segregation, we also simulated disks with two different mass classes ($12.5$ and $25$~M$_\odot$ in a first run, S5, and $10$ and $100$~M$_\odot$ in a second one, S6), while accounting for the appropriate change in the relaxation time, compared with the original model. In both cases we find that the overall evolution of the disk+cluster system is not significantly different from the original S1 model.

Finally, in the S7 run, we tested the dependence on the chosen number of particles by simulating
a disk with $2N_d$ particles (keeping the same total masses of the disks and cluster, but normalizing for the different relaxation time). Again, we find no significant changes compared with our main simulation.
For simplicity, and given the insignificant differences between the various models we tested, we focus here on our main simulation (S1), and show only its results, keeping in mind that they also well represent the results of the other models (see Table \ref{tabICs}). 

\section{Results}\label{sec:res}
If second generation stars form in a disk, it is possible that their 
initial distribution leaves observable signatures in the parent 
dense cluster. Thus, we simulated the evolution of the \omcv-like central disk for a 
period of time that, re-scaled to the mass of the particles, is comparable with the typical
$12$~Gyrs age of Galactic GC. 
 This age corresponds to 0.6~$t_{rh}$, where $t_{rh}$ is the half-mass 
relaxation time of the total system (initially spherical component+disk), evaluated using the system initial properties by mean of Equation (2.63) of \citet{Spi87}.
We then analyzed the evolution of the average angular momentum of the
stars in the disk as well as their orbital parameters in order to find 
relevant kinematic properties that might be detected in real clusters.
Figure \ref{fig:isoc} shows the isodensity contours of the  edge-on disk at the beginning of the simulation and after  0.1, 0.25, and 0.6~$t_{rh}$ (approximately $2$, $5$ and $12$~Gyr after the rescaling). The disk rapidly inflates with time and becomes more spherical but its  radial size is only slightly increased. Although the disk undergoes this fast evolution, its stars remain spatially limited inside $\sim15$~pc, while the tidal radius of the cluster is $\sim 80~$pc.\\
The distribution of the average inclination of the disk stars becomes broader with time, 
and after 0.6~$t_{rh}\sim$12~Gyrs, but already after 0.15~$t_{rh}$, it is much more uniform (see the cumulative distribution evolution in Figure \ref{fig:cosi}), i.e. the disk becomes  more isotropic with time, in agreement with the results in Figure \ref{fig:isoc}.\\
Figure \ref{fig:Lev} illustrates the evolution of the three components of the  average angular momentum (AM) of the  disk  particles (upper panel), showing a consistent decrease with time, and the respective dispersions (bottom panel) that increase with time.
At the end of the simulation the average AM is still significant; even after the long term evolution the disk stars still do not show an isotropic distribution.
Moreover, as the total AM of the system must be conserved, the cluster structure flattens as it exchanges AM with the disk stars. The average component $L_z$ of the angular momentum of particles in the initially spherical component of our system indeed increases with time due to this exchange of AM as it is apparent from Figure \ref{fig:amz}.
\begin{figure}[!t]
\begin{center}
\includegraphics[width=0.4\textwidth]{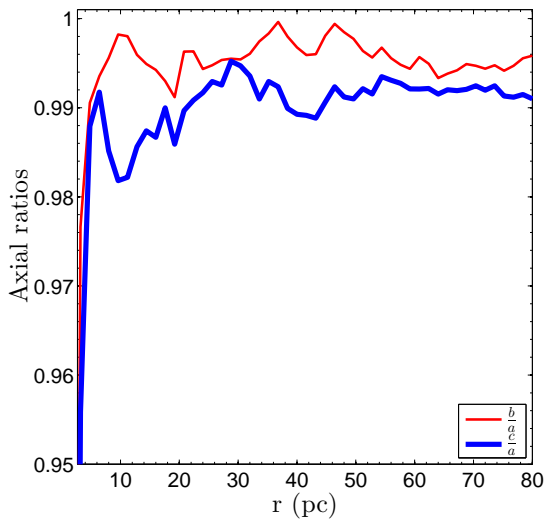}
\caption{ Axial ratios of the initially spherical component of the system at the end of the simulation. The thin (red) line is the axial ratio along the $y$ axis, $b/a$, while the thick (blue) line is the axial ratio along the $z$ axis, $c/a$. }
\label{fig:ax_r}
\end{center}
\end{figure}
\begin{figure*}[!t]
\begin{center}
\includegraphics[width=0.4\textwidth]{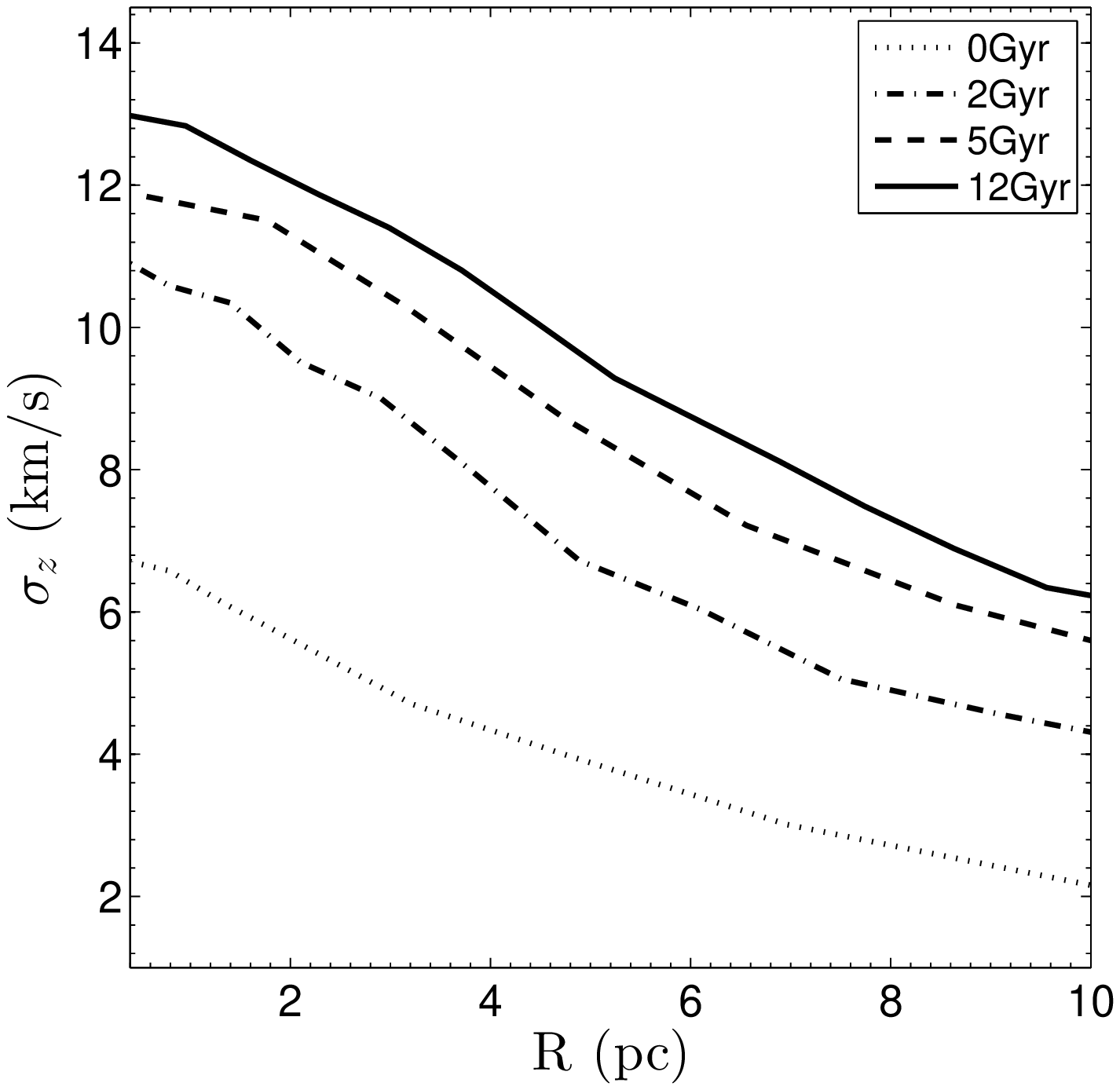}~~~~\includegraphics[width=0.4\textwidth]{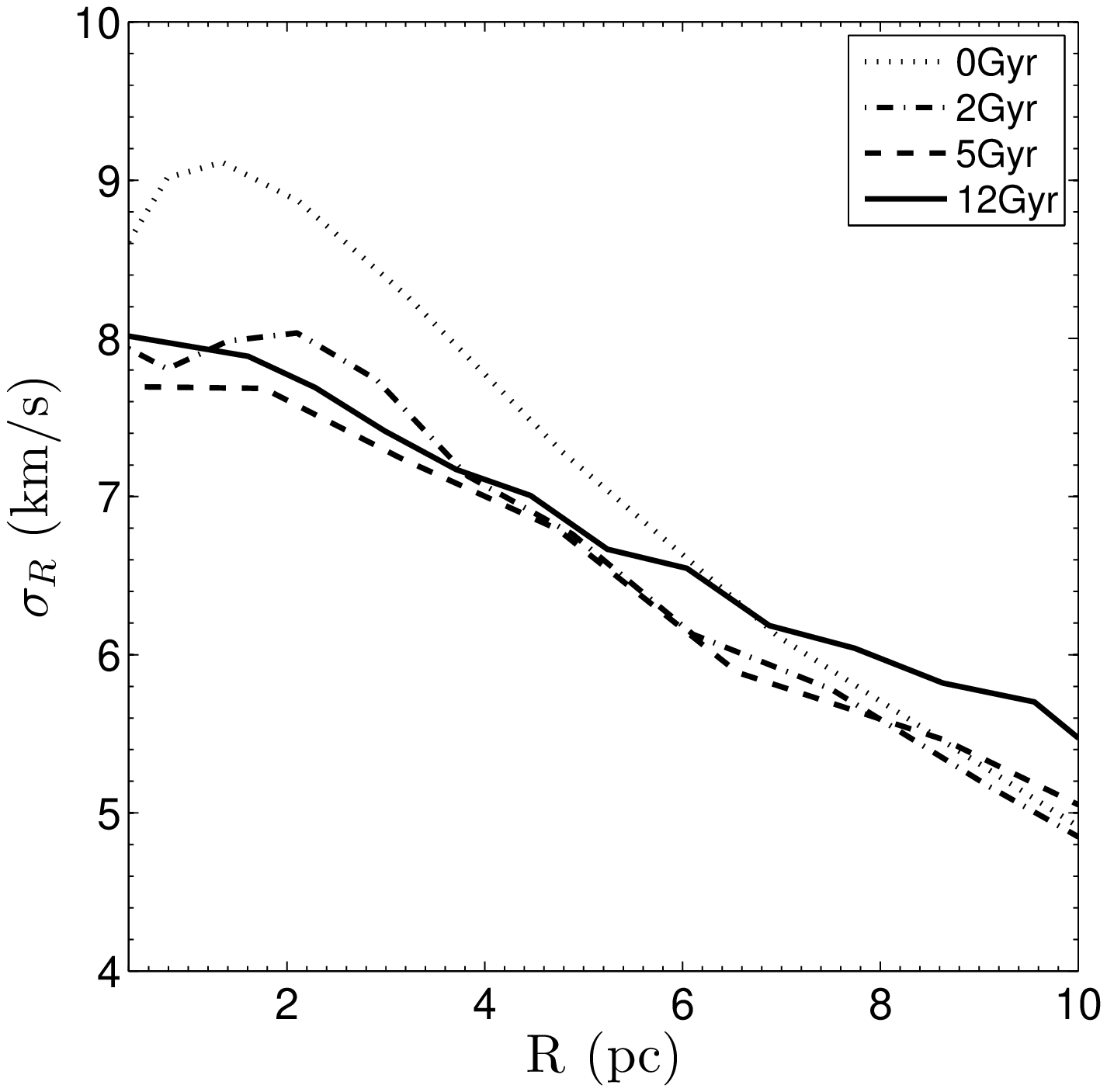}~~~~\\
\includegraphics[width=0.4\textwidth]{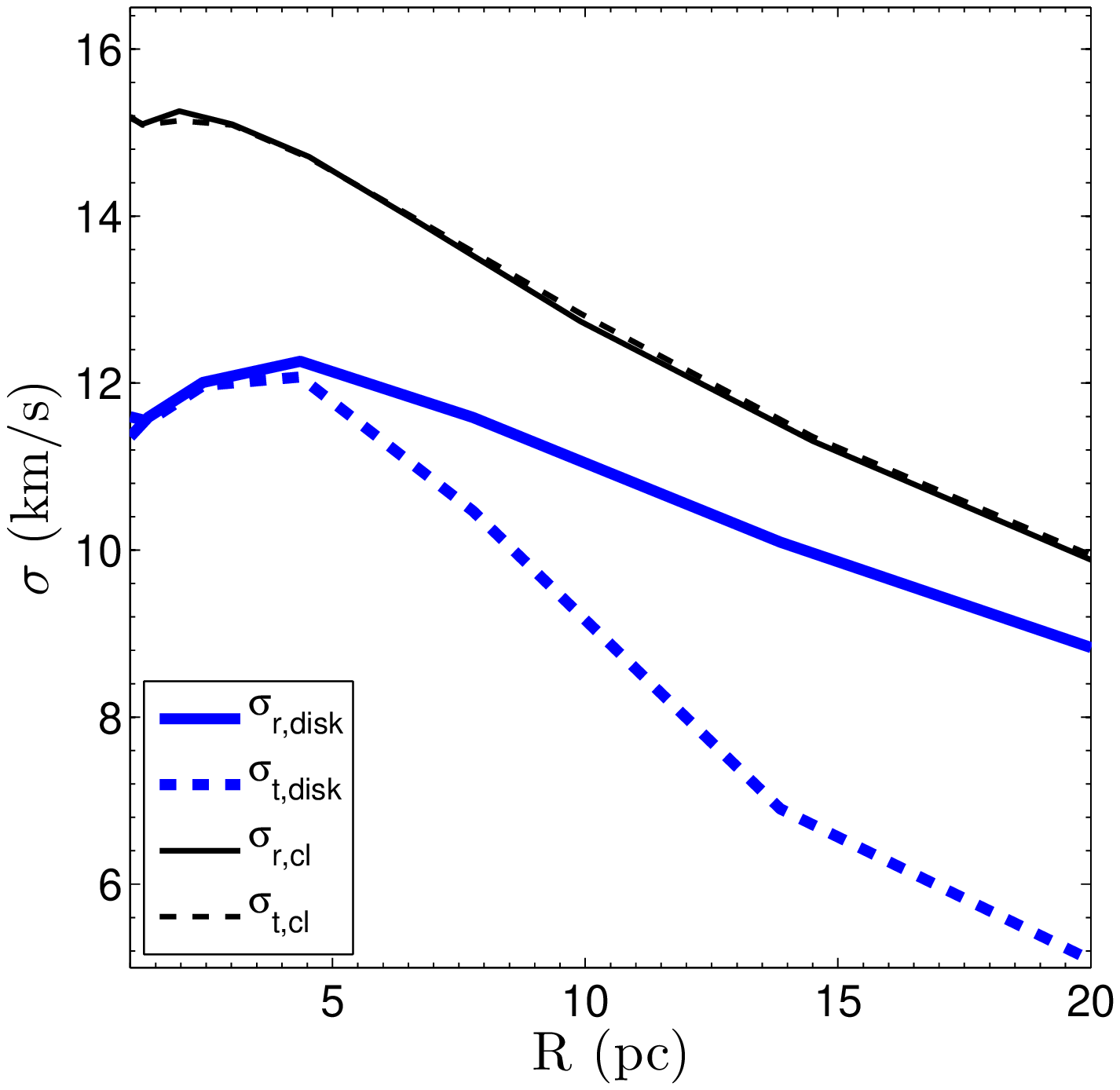} \includegraphics[width=0.41\textwidth]{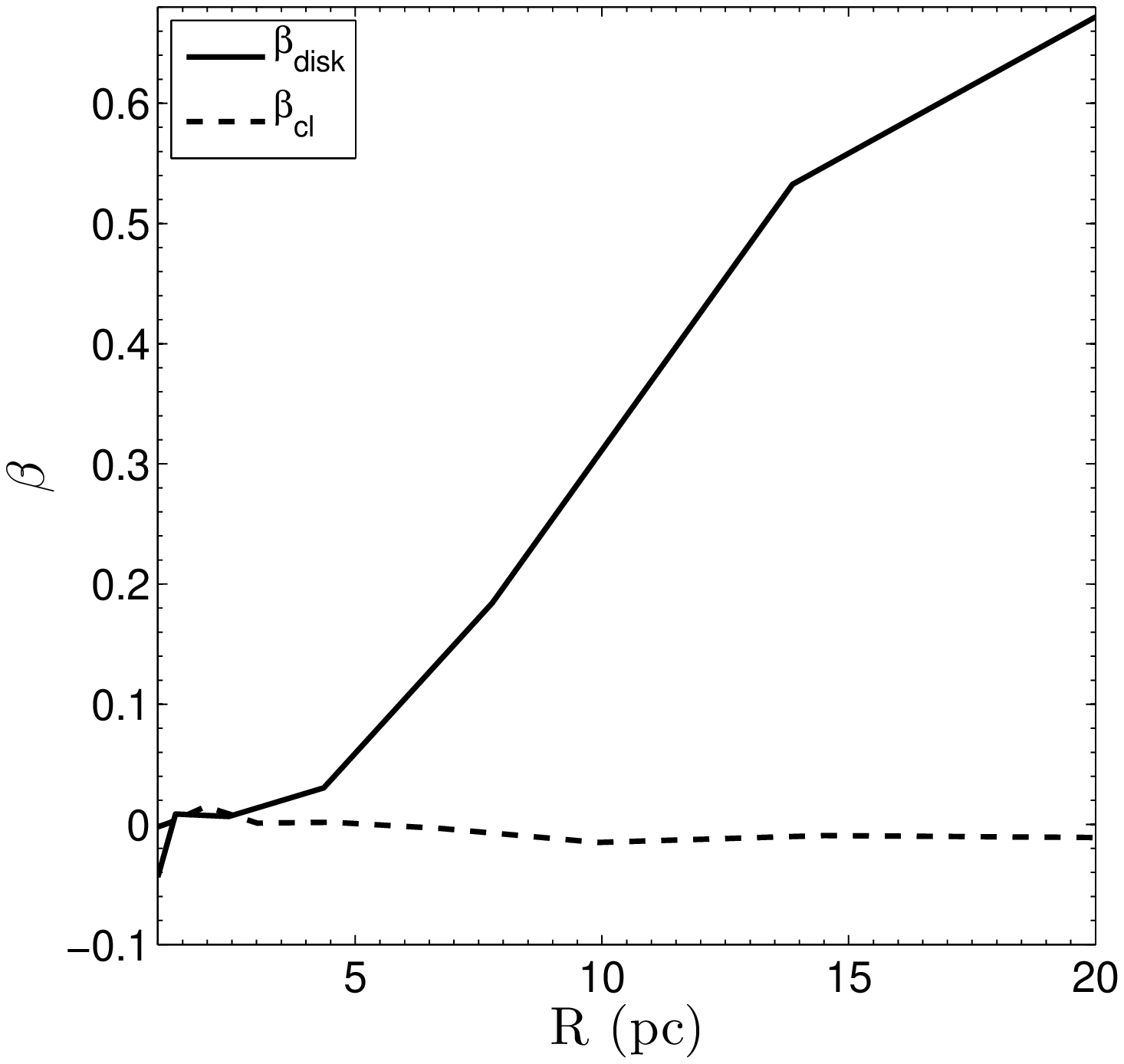}
\caption{Top panels: the velocity dispersion of the disk at four different evolutionary times, 0, 2, 5 and 12~Gyr; $v_z$ is the velocity component perpendicular to the initial plane of the disk while $v_R$ is the radial velocity component on the initial plane of the disk.
Bottom left panel: The final radial (blue thick and black thin solid lines) and the tangential (blue thick and black thin dashed lines) velocity dispersion of the disk and of the cluster as a function of the projected radius. Bottom right panel:
The final anisotropy parameter of the disk (solid line) compared with that of the cluster (dashed line), as a function of the projected radius.  }
\vspace{0.53cm}
\label{fig:iev}
\end{center}
\end{figure*}

\begin{figure}[!t]
\begin{center}
\includegraphics[width=0.405\textwidth]{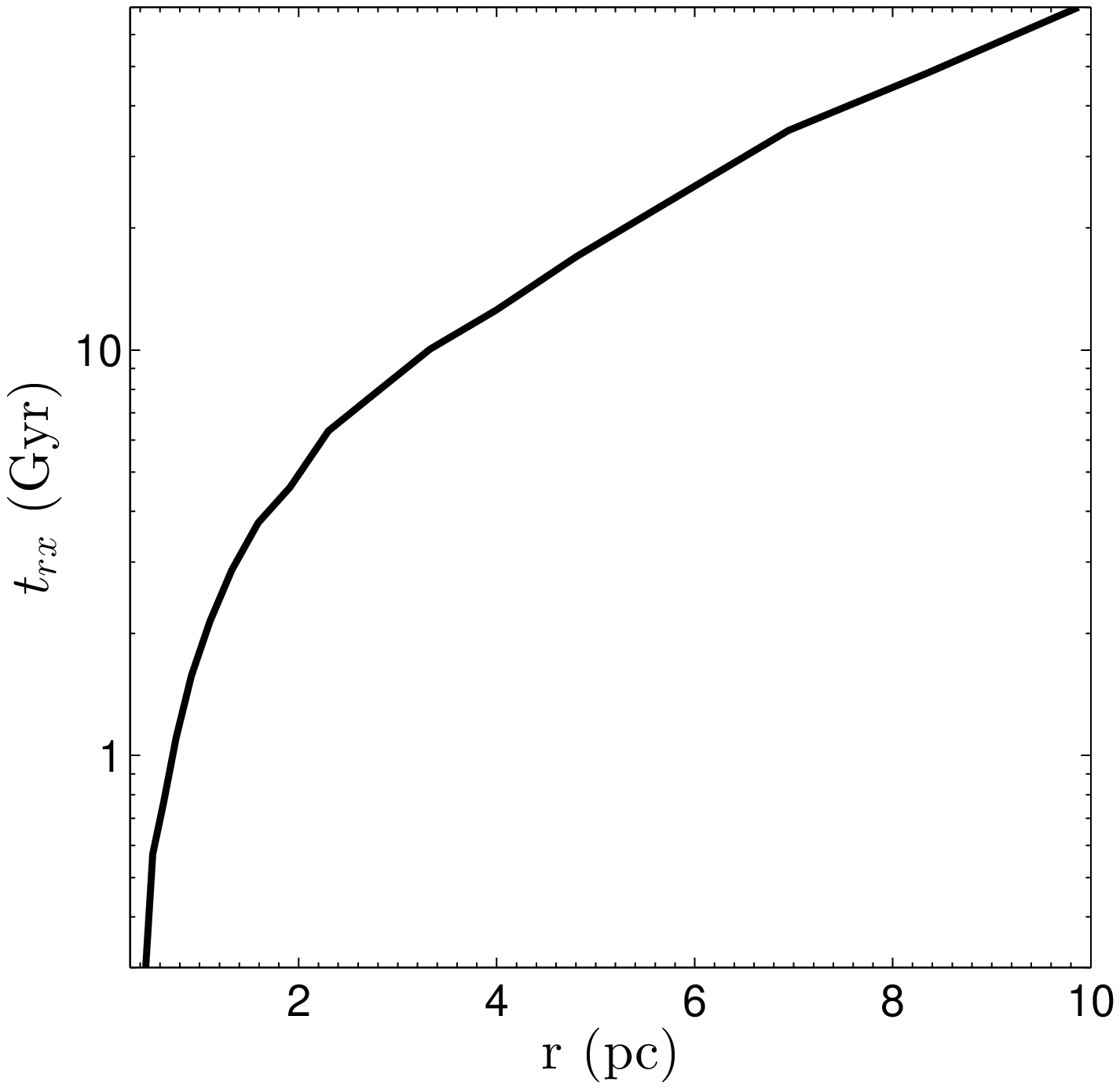}
\caption{ Relaxation time in the central region of the initial simulated system as a function of the radius, obtained using Equation (\ref{eq:relt}). }
\vspace{0.53cm}
\label{fig:trel}
\end{center}
\end{figure}
 To quantify the extent of the flattening we evaluated the axial ratios of the system as a function of the radial distance from the center, by means of the moment 
of the inertia tensor. 
We iterated the procedure described by \citet{K91} until the axial ratios 
inside the spheroid of radius $r^2=x^2/a^2+y^2/b^2+z^2/c^2$ converged with a precision of $5\times 10^{-4}$. 
As can be seen in Figure \ref{fig:ax_r}, the initially spherical cluster component is slightly flattened along the z axis, due to the rotation acquired (see Figure \ref{fig:amz}).  The amount of the flattening
is of $\sim 0.5-5\%$ within $50$~pc from the center of the cluster, while outside this range the system is almost 
axisymmetric.
Interestingly, \citet{VDB08} found a relation between the morphology of the horizontal branch and the flattening of Galactic GCs. It is possible that our results could be tied to this relation, since the shape of the horizontal branch is known to be strongly related to the evolutionary history of the GC and therefore with the presence of MSGs.\\

While the disk inflates, its velocity dispersion in the $z$ direction grows with time, while in the plane of the disk the velocity dispersion
suffers from an initial decrease followed by a smaller increase (see top panels of Figure \ref{fig:iev}). Despite that, at the end of the simulation, the velocity dispersion of the disk stars is still smaller than that of the cluster stars.
Both the radial and the tangential velocity dispersions of the disk and those of the cluster differ, with the former considerably smaller than the latter (see the bottom left panel of Figure \ref{fig:iev}).
Moreover, the disk remains significantly radially anisotropic in  velocity space. Its anisotropy parameter, 
at the end of the simulation, $\beta=1-\sigma_{t}^2/(2\sigma_{r}^2)$ is indeed positive at any radius, 
and has an increasing trend with the distance
from the cluster center (see the bottom right panel of Figure \ref{fig:iev}).

Thus, even after $12$~Gyrs the flat and the spherical components are not yet mixed and can still be spatially and kinematically distinguished. This suggests that signatures of any putative initial distribution of MSGs formed in a disk-like structure might still be observable today \citep[see also][]{Ve13}. This is true at least in clusters (or NCs) with long relaxation times as \omcv, as well as in \omc itself, where many stellar generations have been found.
As a result of the evolution of the disk,  the initially spherical component  becomes slightly tangentially anisotropic with time in the region between $10$ and $20$~pc (see bottom panels of Figure \ref{fig:iev}).
 Outside this radius the cluster becomes radially anisotropic. This trend as been 
also found by \citet{Ta97} in their Fokker-Planck simulations of a cluster with no embedded disk.
Those authors speculate that the presence of the Galactic tidal field should decrease this anisotropy because the external potential would preferentially remove stars on radial orbits from the cluster outskirts. Indeed, in the S4 run, that includes a simplified modeling of the Galactic tides, we observe a decrease of the radial anisotropy in respect to the S1,2 and S3 runs.

\section{Discussion and summary}\label{sec:disc}
GCs are made of several generation of stars whose origin is still
unclear. \omc is the first cluster where this feature has been observed.
Among its peculiarities, this cluster shows a central disk that could be the remnant of the an initial disk where second generation stars formed. 
Indeed \citet{Be11, Be10} showed that the gas ejected by AGB stars can be 
retained inside the parent massive star cluster and can fragment into a nested flattened and rotationally supported system of young stars.
If so, the flattened structure evolves
and may eventually leave kinematic signatures of its former presence.

In this work we employed N-body simulations to explore the long-term evolution of such stellar disks embedded in dense stellar cluster, where for the initial conditions we made use of the observed disk like-structure in \omcv.

We verified that such a disk becomes  rapidly more isotropic with time.
However, even after a time comparable with the age of Galactic GCs, the cluster stars and the second generation disk stars are not yet completely mixed. The two populations have different spatial distributions. The second generation stars are still confined only to the central region of the cluster after 12 Gyrs, and do not attain a completely relaxed spherical shape.
As the second generation disk stars isotropize they exchange angular momentum with the first generation cluster stars, a process which consequently leads to the slight flattening of the host cluster. 
It is worth noting that a more massive disk, as e.g. shown to form in simulations \citep{Be11, Be10} is likely to introduce an even larger flattening.  

We also find that the kinematic properties of the evolved disk stars change with radial distance. The disk preserves a slight ($\beta\sim0.02$) 
radial anisotropy within $\sim 4$~pc; outside this radius $\beta$ grows significantly.
In addition, the second generation stars are also characterized by a significantly smaller velocity dispersion than the first generation stars in the central region, which decreases even further with the distance from the cluster center.
 Thus even though \omcv, in our simulations, seems to 
undergo a fast evolution toward isotropy the signature left both in the spatial distribution and kinematical properties of the disk stars are significant.
 
At this regard, it is interesting to notice that \citet{Ri13} have recently found 
that the bluest and younger stellar population of 47 Tucanae is characterized a by radially anisotropic velocity dispersion. 
Furthermore, the second generation stars in 47 Tucanae appear to be the more centrally concentrated than the reddest and oldest component, in agreement with what we 
found in our test case. 

In cases where relaxation processes are less efficient, initial conditions are better preserved, as the cluster kinematics still hold their dynamical ``memory''. This explains why the kinematics of the disk stars show stronger signatures of their original disk structure as a function of increasing radius, at which relaxation times are longer (see Figure \ref{fig:trel} and Figure 21 in \citealt{VdV06}); indeed, a similar distance dependent effect was suggested to constrain the origin of young stars in the Galactic center \citep{Pe10,Ma13}.
 A clear implication/prediction of this issue is that second generation stellar populations in  GCs and NCs with longer relaxation times, and/or further out from the center, are likely to show stronger anisotropies in their kinematics. Such regions/clusters would serve as the best test sites for identifying fossilized signature of a disk/flattened primordial structure. 

Finally, the observed signatures of such disks may allow us, to some extent,  to determine the dynamical age of the respective stellar population.  Since longer evolved disk stars show weaker kinematic signatures, the level of anisotropy provides a dynamical age stamp for this stellar population (see Figures \ref{fig:isoc}, \ref{fig:cosi} and \ref{fig:iev}), and might even be translated into the real age of the stellar population, after accounting for the host cluster relaxation time. In particular, it it is interesting to note that the current observed disk like configuration in \omcv, suggest a young dynamical age for this stellar component, and may provide clues to its origin.  

Follow up studies exploring larger primordial disk structures than those studied here, beginning with the initial conditions just after second generation star formation (e.g. \citealt{Be11,Be10, MBP13}), will allow us to further explore the generic effects of the long term evolution of second generation stellar disks in a wider range of conditions.  

\acknowledgments{
The authors wish to thank the referee for helpful comments and suggestions. This research was supported by the I-CORE
Program of the Planning and Budgeting Committee and The Israel
Science Foundation grant 1829/12, and in part at the Technion by a fellowship from the Lady Davis Foundation. 
}

\end{document}